# Incorporation of density scaling constraint in density functional design via contrastive representation learning

Weiyi Gong,[1] Tao Sun,[2] Hexin Bai,[3] Shah Tanvir ur Rahman Chowdhury,[1] Peng Chu,[3] Anoj Aryal,[1] Jie Yu,[1] Haibin Ling,[2] * John P. Perdew,[1, 4] * Qimin Yan[1] *

[1]Department of Physics, Temple University, Philadelphia, Pennsylvania 19122, USA

[2]Department of Computer Science, Stony Brook University, Stony Brook, NY 11794, USA

[3]Department of Computer and Information Sciences, Temple University, Philadelphia, Pennsylvania 19122, USA

[4]Department of Chemistry, Temple University, Philadelphia, Pennsylvania 19122, USA

*Correspondence and requests for materials should be addressed to H.L. (haibin.ling@stonybrook.edu), J.P. (perdew@temple.edu), and Q.Y. (qiminyan@temple.edu).

# Abstract

In a data-driven paradigm, machine learning (ML) is the central component for developing accurate and universal exchange-correlation (XC) functionals in density functional theory (DFT). It is well known that XC functionals must satisfy several exact conditions and physical constraints, such as density scaling, spin scaling, and derivative discontinuity. However, these physical constraints are generally not incorporated implicitly into machine learning through model design or pre-processing on large material datasets. In this work, we demonstrate that contrastive learning is a computationally efficient and flexible method to incorporate a physical constraint, especially when the constraint is defined by an equality, in ML-based density functional design. We propose a schematic approach to incorporate the uniform density scaling property of electron density for exchange energies by adopting contrastive representation learning during the pretraining task. The pretrained hidden representation is transferred to the downstream task to predict the exchange energies calculated by DFT. Based on the computed electron density and exchange energies of around 10,000 molecules in the QM9 database, the augmented molecular density dataset is generated using the density scaling property of exchange energy functionals based on the chosen scaling factors. The electron density encoder transferred from the pretraining task based on contrastive learning predicts exchange energies that satisfy the scaling property, while the model trained without using contrastive learning gives poor predictions for the scaling-transformed electron density systems. Furthermore, the model with pretrained encoder gives satisfactory performance with only small fractions of the whole augmented dataset labeled, comparable to the model trained from scratch using the whole dataset. The results demonstrate that incorporating exact constraints through contrastive learning can enhance the understanding of density-energy mapping using neural network (NN) models with less data labeling, which will be beneficial to generalize the application of NN-based XC functionals in a wide range of scenarios which are not always available experimentally but theoretically justified. This work represents a viable pathway toward the machine learning design of a universal density functional via representation learning.

# Introduction

Density functional theory (DFT) is an indispensable tool in computational chemistry and materials science due to its combination of efficiency and accuracy.[1, 2] As the standard computational method that is widely applied in physics, chemistry, and materials research, DFT has achieved high prediction accuracy enabled by the continued development of approximations of the exchange-correlation (XC) energy as a functional of electron density.[3-7] An appropriately approximated density functional enables more accurate first-principles calculations for molecules and material systems on a larger scale. In different forms of approximations, the XC functionals must satisfy several exact conditions and constraints, such as uniform scaling property,[8] spin scaling property[9] and derivative discontinuity.[10] So far, all popular approximations suffer from systematic errors that arise from the violation of mathematical properties of the exact functional. It is expected that the performance and generality of density functionals can be improved by satisfying these constraints. For instance, the recently developed strongly constrained and appropriately normed (SCAN) functional[7] that satisfied 17 exact constraints achieved great performance for both molecules and solids.

There has been a growing interest in applying machine-learning (ML) in physics, chemistry, and material science, with the aim of achieving the same or even higher prediction accuracy for molecules and materials with much less computational cost compared to first principles simulations. Recently, ML has been applied to parametrize XC functionals without domain knowledge of humans by using various methods such as kernel ridge regression (KRR),[11] fully connected neural networks (NN)[12-14] and convolutional neural networks (CNN).[15] Being trained in a supervised manner, these ML models are highly accurate across a small set of molecule systems similar to those on which the models are trained, while in many cases they show a worse performance on larger molecular datasets than they do on small ones. Neither of them demonstrates the same level of universality compared to conventional XC functionals.

Plenty of effort has been devoted to leveraging physical constraints in ML of XC functionals. In a previous work by Lei *et al.*,[15] by using CNN as encoders, rotationally invariant descriptors were extracted and projected on a basis using spherical harmonic kernels. In another work by Hollingsworth *et al.*,[16] it was found that the scaling property, which is one of the exact conditions that the exchange energy must satisfy, can be utilized to improve the machine learning of XC functionals. The study is limited to one-dimensional systems and lacks the generalizability to two- and three-dimensional systems. More recently, another exact condition - derivative discontinuity - was incorporated into the NN-based XC functional design,[17] while the study is again limited to one-dimensional systems. A more recent work has demonstrated that the fundamental limitation can be overcome by training a neural network on molecular data and on fictitious systems with fractional charge and spin,[18] and the resulting NN-based functional DeepMind 21 demonstrated the universality and greatly improved predictive power for molecule energetics and dynamics.

Many of the previous works use data augmentation to improve model performance by directly increasing the amount of labeled data following a given physical constraint. However, increasing the amount of data is not always possible due to the computational cost. Going beyond data augmentation, self-supervised learning has gained popularity because of its ability to avoid the cost of annotating large-scale datasets. It adopts self-defined pseudo labels as supervision and uses the learned representations for downstream tasks. Self-supervised learning has been widely used in image representation learning[19] and natural language processing,[20] and has been applied in molecular machine learning.[21, 22] Specifically, contrastive learning (CL) has recently become a dominant branch in self-supervised learning methods for computer vision, natural language processing, and other domains.[23] It aims at embedding augmented versions of the same sample close to each other while trying to push away embeddings from different samples in the representation space. The goal of contrastive learning is to learn such an embedding space in which similar sample pairs stay close to each other while dissimilar ones are far apart, and the CL process can be applied in both unsupervised and supervised settings.[24] In this work, we will explore the incorporation of physical constraints in density functional learning through contrastive learning.

One of the most important and fundamental constraints for the exchange energy of an electron system is derived from the principle of uniform scaling.[8] Consider an electron density distribution $n(r)$ and a scaled density $n_\gamma(r) = \gamma^3 n(\gamma r)$ (where $\gamma$ is a scaling factor), several important exact constraints on exchange and correlation energy functionals can be written using the scaled density. In this work, we focus on the exchange energy $E_x[n]$, and the scaling property must satisfy the following requirements: $E_x[n_\gamma] = \gamma E_x[n]$. This important constraint is satisfied exactly in almost all human-designed density functionals, whether non-empirical or semi-empirical. As a chemical example, atomic one-electron ions of nuclear charge $Z$ are scaled versions of. the hydrogen atom with scale factor $\gamma = Z$. The exchange energy, $-5Ze^2/(16a_0)$, in this case cancels the Hartree electrostatic interaction of the density with itself. Using this constraint as an outstanding example, we propose a schematic approach to incorporate any physical constraints (represented by equalities) via contrastive learning into the NN-based model design.

Specifically, we pre-trained an electron density encoder by maximizing the similarity between molecular electron density and its scaled version with a randomly chosen scaling factor, within the framework of SimCLR[25], which is a widely used framework for contrastive learning of image pretraining. To obtain an encoder that gives similar representations (while different by a scaling factor) for scaled and unscaled electron densities, we added a scaling factor predictor component to the framework. The pre-trained encoder was then transferred to the downstream task to predict the exchange energies from electron densities of molecule systems. It is shown that the model pretrained contrastively predicts exchange energies that satisfy the scaling relation, while the model trained without using contrastive learning gives poor predictions. We will show that contrastively learned encoders are capable of encoding molecular electron density with less labeling cost based on the fact that they give comparable predictions by fine-tuning using only a small percentage of labeled data, compared to the model trained on the whole labeled dataset by supervised learning. This shows that contrastive learning using constraints can enhance the understanding of DFT theory for neural network models with a small amount of labeled data while generalizing the application of NN XC functionals in a wide range of scenarios which are not always available experimentally but theoretically justified.

## Results

**Electron density encoder**

In this work, to efficiently handle a large amount of three-dimensional grid-based electron density data, the Residual Network (ResNet) was used as the electron density encoder. ResNet is one of the most commonly used networks in image recognition. With deeper and deeper neural networks, effective learning becomes more challenging due to the gradient vanishing or exploding problem,[26,27] which makes traditional models using convolutional neural network layers reach a limit of performance when the number of layers increases. In 2016, He *et al.*[28] proposed using skip-connection that allows direct connection from the input layer to the output. By skipping intermediate layers, the model is able to learn the identity map even if there is a gradient issue within these layers. Instead of learning the mapping $H$ between input $x$ and target $y$, residual networks aim to learn the residual $F$:

$$F(x) := H(x) - x$$

In the worst case, a trivial result is learned such that $F(x) = 0$, the mapping $H$ is the identity mapping $H(x) = x$. This skip-connection architecture enables the learning ability of neural networks that are extremely deep, which is critical for large-scale three-dimensional electron densities.

**Contrastive learning of uniform density scaling property**

Contrastive learning (CL) is a self-supervised learning (SSL) strategy that learns useful representations using unlabeled data by manually designing pre-training tasks with automatically generated labels or label relations. Typically, when applied in image recognition, data augmentations such as random shifting, random cropping and random rotation are applied to generate different views of images. The raw and augmented images are then passed to an image encoder to generate hidden representations that are passed to a projection head projecting representations onto a high dimensional unit sphere. The projected representations are used to calculate contrastive loss that maximizes the similarity between projected representations of the same input image, while minimizing the similarity between those of different images. By minimizing contrastive loss and updating the model parameters through backpropagation, the image encoder is aware that the different views are from the same raw image, which introduces invariance to the model for imperfect inputs. Intuitively, an encoder trained by contrastive learning groups different views of the same image into the same cluster while pushing clusters from different images far away from each other.

In this work, we intend to design a pre-training task such that the electron density encoder is aware of the uniform density scaling property. In order to do so, unscaled and scaled electron densities on a fixed-size spatial grid are generated using the PySCF code[29] with low computation cost, represented as three-dimensional arrays $x_i, \tilde{x}_{i\gamma} \in \mathbb{R}^{d \times d \times d}$, where the scaling factor $\gamma$ is chosen from four different scales: 1/3, 1/2, 2, and 3. Electron density arrays are encoded as hidden representations $h_i = f(x_i), \tilde{h}_{i\gamma} = f(\tilde{x}_{i\gamma}) \in \mathbb{R}^m$ through the density encoder that is a mapping $f: \mathbb{R}^{d \times d \times d} \to \mathbb{R}^m$ to be learned. The hidden representations are then projected as a set of points $z_i = g(h_i) \in \mathbb{R}^n$ on a high dimensional unit sphere by a mapping $g: \mathbb{R}^m \to \mathbb{R}^n$ ($n < m$) that is a multilayer perceptron (MLP). For a batch of $N$ molecules, the output $Z \in \mathbb{R}^{2N \times m}$ contains projected representations of unscaled and scaled densities. Then we calculate the normalized temperature-scaled cross entropy (NT-Xent) loss[25] that is defined as:

$$l_{ij} = -\log \frac{\exp(z_i \cdot z_j / \tau)}{\sum_{k=1, k \neq i}^{2N} \exp(z_i \cdot z_k / \tau)},$$

where the temperature factor $\tau$ is a small positive real number, and the exponential term when $k = i$ is excluded in the summation in the denominator to ensure that the loss is zero if dissimilar projected representations are antiparallel. Indeed, for $\tau \to 0^+$, $z_i \cdot z_j \neq 0$, $z_i \cdot z_k = -1$ ($k \neq j$),

$$l_{ij} = \log\left(1 + \frac{\sum_{k=1, k \neq i,j}^{2N} \exp(z_i \cdot z_k / \tau)}{\exp(z_i \cdot z_j / \tau)}\right) = \log\left[1 + (2N - 2)\exp\left(-\frac{2}{\tau}\right)\right] \to 0$$

For a batch of $N$ molecules, $z_{2k-1}$ and $z_{2k}$ are the corresponding projected representations of unscaled and scaled densities of the same molecule. Notice that the loss function is asymmetric ($l_{ij} \neq l_{ji}$) and the total loss is

$$L = \frac{1}{2N} \sum_{k=1}^{N} (l_{2k-1,2k} + l_{2k,2k-1})$$

The loss is zero when the projected representations of different molecules are perpendicular to each other, which ensures that dissimilar samples are pushed far apart from each other.

In the original SimCLR framework[25], augmented and unaugmented views of the same input form positive pairs, while those of different inputs form negative pairs. We would emphasize that, without any modules added to distinguish positive pairs, the encoder trained would be too "lazy" to learn different representations for the two "views" of the same input, since the simplest mapping $f$ that minimize the loss learns the same hidden representation for the augmented and unaugmented input from the same image, which satisfies $\tilde{h}_{i\gamma} = f(\tilde{x}_{i\gamma}) = f(x_i) = h_i$. Therefore, a module predicting the scaling factor from two hidden representations of the same molecule is added to distinguish the scaled density data from unscaled ones. The final loss of the contrastive pretraining task is the summation of these two losses. The workflow of the pretraining task is shown in Fig. 2(a).

The cosine similarity of learned projected representations $z$ and $\tilde{z}$ for a batch of 32 molecules are shown in Fig. 3(a). As expected, the cosine similarity shows maximum values for positive pairs – unscaled and scaled densities of the same molecules, while the value is close to zero for negative pairs – densities of different molecules. We further verify that projected representations of different molecules are well separated from each other by computing the t-distributed neighbor embedding (t-SNE). In Fig. 3(c), two examples of molecules, learned projected representations and predictions on scaling factors are shown. The best model achieves 0.01976 contrastive loss and 2e-4 mean square error for scaling factor prediction.

**Comparison of performance of supervised learning and contrastive learning**

Supervised learning of neural networks is one of the most widely used machine learning strategies in material science. By training with a large amount of labeled data, which means that each input has the corresponding target, the model can give predictions with a small discrepancy with the true targets. However, one of the limitations of supervised learning is the fact that an outstanding performance on a given dataset does not guarantee equally good performance on other datasets. In this section, we will show that the model trained by supervised learning on unscaled density data achieves a very high prediction accuracy for predicting exchange energies from unscaled molecular electron densities, but at the same time demonstrates a large prediction error for scaled densities. This observation clearly shows that the model trained on unscaled density dataset with supervised learning does not understand the uniform scaling property that exchange energy functionals must satisfy.

Within the data-driven paradigm, the mapping of molecular electron density to the exchange energy is directly learned in a supervised manner by feeding electron densities to an electron density encoder, with the corresponding exchange energies calculated from first-principles calculations as labels. Electron density in three-dimensional space is represented by a three-dimensional array, with the dimension along each axis equal to the grid dimension along the same axis. Encoding and decoding of volumetric data in three-dimensional space has been previously studied in 3D-UNet,[30] with a DoubleConv layer consisting of two subsequent 3D convolutional layers as the building block. In the same 3D-UNet framework, instead of DoubleConv, residual networks can be used as the building block to extract useful information from raw three-dimensional volumetric data.[31] In this work, the mapping of electron density to the exchange energy will be learned, so only the encoder part will be adopted from 3D-UNet. The encoder consists of several connected building block layers, being either DoubleConv or ResNet (see Methods).

The architecture of the encoder is shown in Fig. 1(b). A hidden representation that captures density-energy correlation is learned and fed to a subsequent fully connected prediction layer to give a single value prediction on the exchange energy. The original electron densities of molecules (with a scaling factor equal to one) are included in the dataset. For reliable evaluation of the models, the dataset is split into 80% and 20% as training and validating datasets. The training set is employed to train the model for 500 epochs by minimizing the mean squared error (MSE) loss, and the model is then applied to validate the performance on the validation set using the mean absolute error (MAE) as the measure.

To investigate whether the model trained with only unscaled densities understands the uniform density scaling property, we test its performance on both unscaled and scaled density datasets. As shown in Fig. 3(a), the difference in energy between predictions and targets on the unscaled dataset is close to 0.45 eV on average. Instead of minimizing this prediction error for unscaled electron density by improving existing learning frameworks, the focus in this work is to demonstrate the role of contrastive learning in the process of incorporating physical constraints in density functional design. As shown in Fig. 3(a), a clear observation is that the model does not provide reasonable predictions for the exchange energies of scaled density dataset. This indicates that the models trained in a supervised manner in general do not satisfy the uniform density scaling property and thus give unreliable predictions for scaled densities, although they may achieve very high accuracy on the unscaled density dataset. This motivates us to apply contrastive learning in a pretraining task to give our model the ability to understand the density scaling property.

**Contrastive learning model performance with different label percentages**

Now we investigate the model for predicting exchange energies from electron densities. The density encoder part of the model is transferred from the contrastive pretraining task. In a comparative test, the model is trained from scratch and its performance is compared to the transferred model. When fine-tuning the transferred model, we adopt training sets with 80%, 60%, 40%, or 20% labeled data. Surprisingly, the transferred model outperforms the model trained from scratch with label percentage as low as 60%. This demonstrates that our contrastive learning model can reduce the need for a large amount of data while achieving even better performance.

Furthermore, the model trained with the contrastive learning method gives a prediction of exchange energies that satisfy the uniform density scaling property. As shown in Fig. 4, predicted and target exchange energies demonstrate a strong linear correlation even when the label percentage is decreased. Note that for the case of 20% label percentage, the model uses the same number of labels as that of the supervised learning task in a previous section. The dramatic difference of performance between models shown in Fig. 3 shows the understandability of uniform scaling property which is enabled by our proposed models.

## Discussion

In this work, contrastive learning is adopted to a pretrained electron density encoder to incorporate the uniform density scaling property for exchange energy predictions. Generated from first-principles calculations, the scaled and unscaled electron densities of molecules from the QM9 dataset are used to contrastively train the electron density encoder. Scaled and unscaled densities of the same molecule are treated as similar pairs, while those from different molecules as dissimilar ones. The pretrained model achieves a 0.01976 contrastive loss. It also predicts the scaling factors from hidden representations of scaled and unscaled densities, with a 2e-4 MSE accuracy. The encoder is then transferred to a downstream task to predict the computed exchange energies from electron densities with different scaling factors. Using contrastive learning as the pretraining method, our model performs well for the prediction of exchange energies of both scaled and unscaled electron densities that satisfy the uniform scaling property, while the model trained using only unscaled densities in a supervised manner demonstrates unreliable performance for the prediction of exchange energies of scaled densities. This clearly demonstrates that contrastive learning is an effective approach in a data-driven paradigm to enable the neural network to learn physical principles in the process of mapping electron densities to energies.

We show that contrastive learning can be used as an adaptive and effective method to incorporate the uniform scaling property of DFT theory into the machine learning model design. Moreover, the contrastive learning method proposed in this work has the potential to be generalized to other exact physical constraints, such as rotational symmetry, spin scaling property, and so on. From this point of view, incorporating physical constraints into machine learning model design through contrastive learning can lead to a significant reduction of the need of training data while providing insights into the machine learning XC density functionals and beyond.

A similar effect occurs with human-designed density functionals: Those that are constructed to satisfy more exact constraints require fewer fit parameters that can be determined from smaller sets of molecular data, and a nonempirical meta-GGA functional[7] satisfying 17 exact constraints can perform rather well without any fitting to molecular data. The improvement of generalized gradient approximations (GGAs) or meta-GGAs by their global hybridization[32] with exact exchange is a good example, since the exact constraints on the underlying GGA or meta-GGA are preserved for any value of the fraction of exact exchange that is mixed with a complementary fraction of GGA or meta-GGA exchange.

## Methods

**Molecular electron density dataset**

We chose 10k molecules from the QM9 dataset [33, 34] by imposing the following criteria: (i) each molecule contains less than 20 atoms; (ii) each molecule does not contain atoms with an atomic number larger than 36 (element Kr); (iii) the size of each molecule is less than 12 angstroms; and (iv) the DFT calculated exchange energy of the molecule should be greater than -200 eV. Molecular density matrices are calculated by DFT with the PBE functional[3] as implemented in the PySCF package.[29] To prepare the grid-like input data with fixed dimensions, we project the density matrices onto real space grid points with a shape (65, 65, 65) on a fixed size cube centered at the origin with a length of 40 angstroms. The number of grid points is set to odd integers to include the origin. A larger grid with shape (129, 129, 129) is also used to construct more detailed density data. Due to the limit of storage for the whole dataset, an average pooling down-sampling pre-process is applied to reduce the grid dimensions from 129 to 65. A comparison of the results using these two grids is given in later sections. The projection of density matrices on grids in three-dimensional space is performed by using the PySCF code.[29] The exchange energies are calculated from the density matrices as they would be in Hartree-Fock or exact exchange theories using the NWChem code[35].

**Training and evaluation of supervised learning task**

To find out the best model that encodes the electron density, two different types of building block layers: *ResNet* and *DoubleConv,* were used to build the density encoder. The model was built and trained using the Pytorch-Lightning package[36] which is a framework based on the Pytorch package[37]. The whole dataset is split into 80% and 20% for training and validating, respectively. Training loss is backpropagated to update the model parameters by an Adam optimizer[38] with a learning rate of 0.001. The best model was chosen to be that with smallest MAE after 500 epochs.

## Acknowledgments

W. Gong and Q. Yan acknowledge support from the U.S. Department of Energy, Office of Science, under award number DE-SC0020310. S.T.U.R. Chowdhury and J.P. Perdew acknowledge support from the U.S. National Science Foundation under Grant No. DMR-1939528. This work benefitted from the supercomputing resources of the National Energy Research Scientific Computing Center (NERSC), a U.S. Department of Energy Office of Science User Facility operated under Contract No. DE-AC02-05CH11231. H. Ling acknowledge the support from SBU-BNL Seed Grant.

## Competing Interests

The authors declare that they have no competing interests.

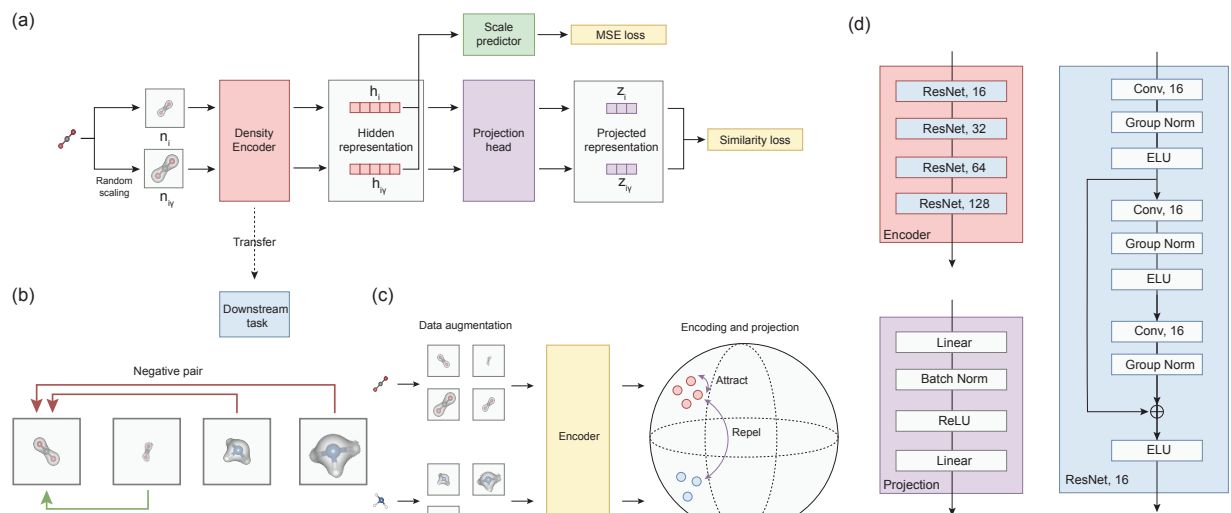

Figure 1. (a) The workflow of the proposed contrastive learning framework. For a given molecule, an unscaled and a scaled electron density are fed into the density encoder to obtain hidden representations. The subsequent modules are divided into two parts: a projection head that produces the projected representations, from which the contrastive similarity loss is calculated; a scale predictor that predicts the scaling factor from the hidden representation pairs, from which the mean squared error loss is calculated. (b) The two electron densities from the same molecule form positive pairs, while those from different molecules form negative pairs. (c) The visualization of general contrastive learning. Multiple "views" of the same input molecule are generated by data augmentation. After encoding and projection, representations from the same molecule attract each other, while those from different molecules repel each other. (d) The architecture of the density encoder, the projection module, and the ResNet building block.

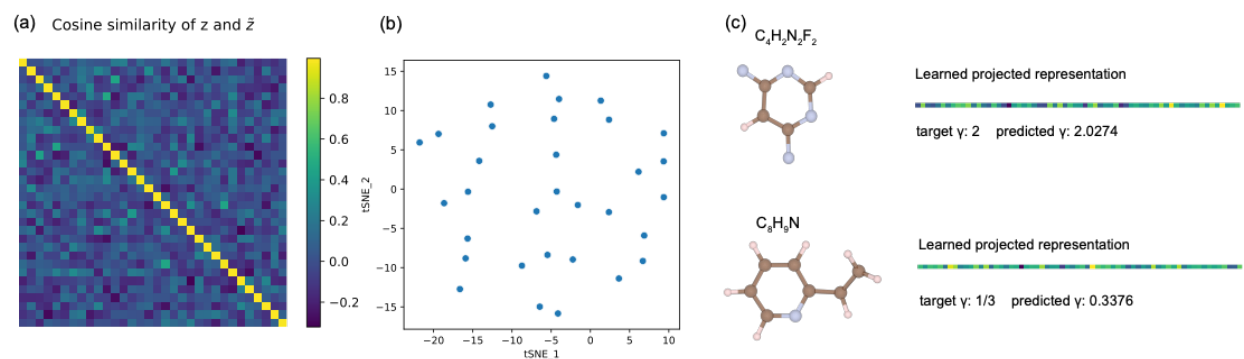

Figure 2. (a)The cosine similarity between the learned projected representations of unscaled and scaled densities for a batch of 32 molecules. Each element in the matrix is computed as $cos(z_i, \tilde{z}_j) := z_i \cdot \tilde{z}_j$. The brighter it is, the closer the value is to 1. (b) The t-distributed stochastic neighbor embedding (t-SNE) of 32 learned projected representations. (c) Two molecule examples, the corresponding learned projected representations, and the predictions on scaling factors.

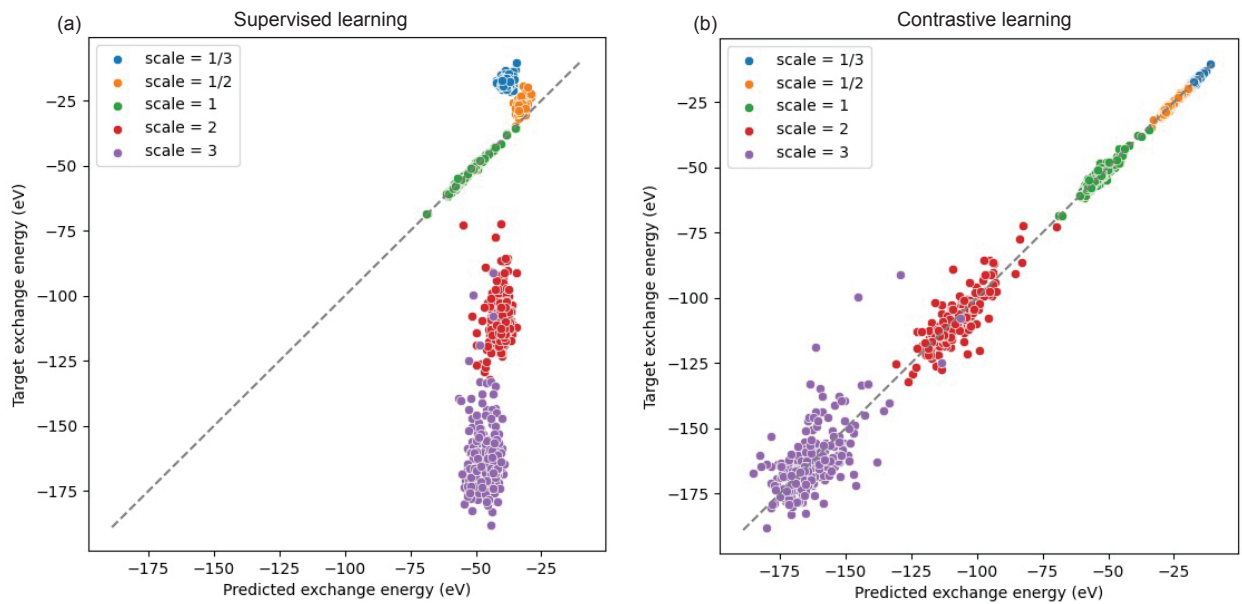

Figure 3. Comparison of performance between supervised learning model and contrastive learning model on datasets with different scaling factors. (a) Model trained by supervised learning on unscaled dataset (green) gives poor predictions for scaled datasets. (b) Model trained by contrastive learning give much more reliable predictions on all datasets (both scaled and unscaled).

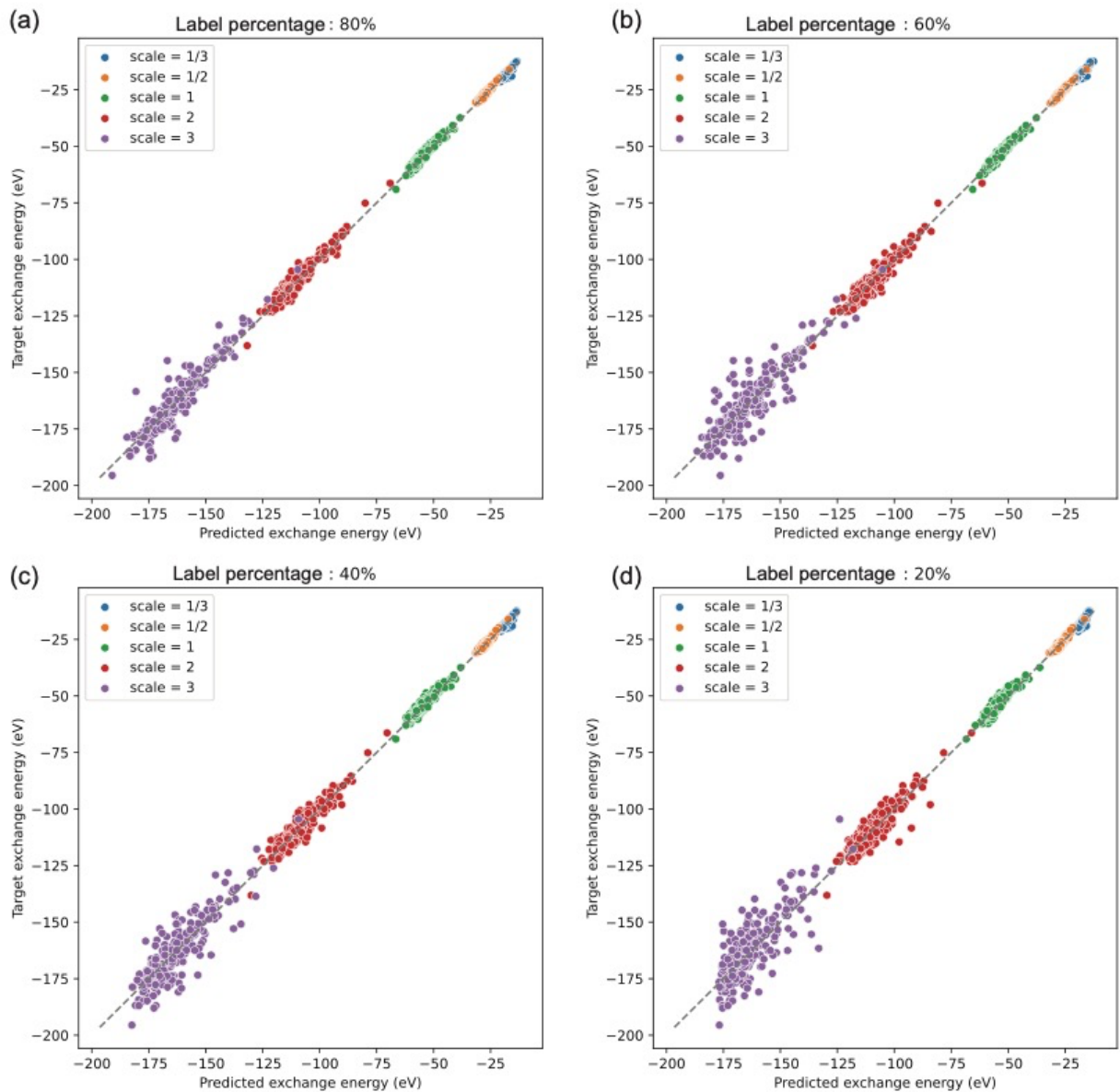

Figure 4. Prediction vs target for four models fine-tuned with four different label percentages. The model keeps the capability to give relatively good predictions even when the label percentage is decreased.

Table 1. The MAE of model (in eV) with ResNet and DoubleConv as density encoders for predicting exchange energies of molecule systems in the QM9 database. Performance is tested for model trained from scratch in a supervised manner and models trained in a contrastive learning plus transfer learning scheme with 80%, 60%, 40%, and 20% labeled data.

| Label percentage | ResNet (16, 32, 64, 128) | DoubleConv (32, 64, 128) |
|---|---|---|
| Supervised learning | | |
| 100% | 2.2673 | 2.6107 |
| Contrastive + transfer learning | | |
| 80% | 1.8265 | 2.2000 |
| 60% | 2.0353 | 2.2319 |
| 40% | 2.3249 | 2.5910 |
| 20% | 2.5472 | 2.9377 |